\documentclass[preprint,12pt]{emulateapj}
\usepackage{multirow}
\newcommand{\water}{H$_2$O}
\newcommand{\methane}{CH$_4$}

\providecommand{\eprint}[1]{\href{http://arxiv.org/abs/#1}{#1}}
\providecommand{\adsurl}[1]{\href{#1}{ADS}}
\def\apss{{Ap\&SS}}		% Astronomy and Astrophysics
\def\aap{{A\&A}}		% Astronomy and Astrophysics
\def\apj{{ApJ}}			% Astrophysical Journal
\def\apjl{{ApJ}}		% Astrophysical Journal, Letters
\def\apjs{{ApJS}}		% Astrophysical Journal, Supplement
\def\pasp{{PASP}}		% Publications of the ASP
\def\procspie{{Proc.~SPIE}}
\def\pasa{{PASA}}
\def\mnras{{MNRAS}}
\def\nat{{Nature}}

\shorttitle{Exoplanets and ELTs}
\shortauthors{Crossfield}

\begin{document}

\title{Exoplanet Atmospheres and Giant Ground-Based Telescopes}

\author{
Ian J. M. Crossfield\altaffilmark{1}$^,$\altaffilmark{2}
}
\altaffiltext{1}{Lunar \& Planetary Laboratory, University of Arizona, 1629 E. University Blvd., Tucson, AZ, USA, \href{mailto:ianc@lpl.arizona.edu}{ianc@lpl.arizona.edu}}
\altaffiltext{2}{NASA Sagan Fellow}

%% \author{\speaker{Ian J.\ M.\ Crossfield}\thanks{A footnote may follow.}\\
%%         NASA Sagan Fellow.\\
%%         University of Arizona, Lunar and Planetary Laboratory.\\
%%         E-mail: \email{ianc@lpl.arizona.edu}}

%\author{Another Author\\
%        Affiliation\\
%        E-mail: \email{...}}

\begin{abstract}The study of extrasolar planets has rapidly expanded to
  encompass the search for new planets, measurements of sizes and
  masses, models of planetary interiors, planetary demographics and
  occurrence frequencies, the characterization of planetary orbits and
  dynamics, and studies of these worlds' complex atmospheres.  Our
  insights into exoplanets dramatically advance whenever improved
  tools and techniques become available, and surely the largest tools
  now being planned are the optical/infrared Extremely Large
  Telescopes (ELTs).  Two themes summarize the advantages of atmospheric
  studies with the ELTs: {\bf high angular resolution} when operating
  at the diffraction limit and {\bf high spectral resolution} enabled
  by the unprecedented collecting area of these large telescopes.
  This brief review describes new opportunities afforded by the ELTs
  to study the composition, structure, dynamics, and evolution of
  these planets' atmospheres, while specifically focusing on some of
  the most compelling atmospheric science cases for four qualitatively
  different planet populations: highly irradiated gas giants, young,
  hot giant planets, old, cold gas giants, and small planets and Earth
  analogs.
\end{abstract}

%% \FullConference{BASH 2015\\
%% 		18 - 20 October, 2015\\
%% 		The University of Texas at Austin, USA}

%\begin{document}

\section{Introduction and Overview}
Over the past two decades, the study of extrasolar planets has 
grown more rapidly than any other field of astronomy.  Once the
province of only a small number of explorers, hundreds of researchers
across our globe now work to find new planets, measure their sizes and
masses, model planetary interiors, measure the intrinsic frequency
with which planets occur, characterize planetary orbits and dynamical
interactions, and observe and model the complex atmospheres of these
other worlds.

Our insights into exoplanets dramatically advance whenever improved
instruments, facilities, and/or observing techniques become available.
Surely the largest astronomical assets now being planned are the
so-called Extremely Large Telescopes (ELTs) -- the next generation of
large-aperture, optical/infrared-optimized, ground-based telescopes.
These include the 25\,m-diameter GMT \citep{johns:2008}, 30\,m TMT
\citep{nelson:2008}, and the 39\,m E-ELT \citep{gilmozzi:2007}. With so
many resources already committed to these projects, it seems likely
that at least one ELT (and hopefully more) will see first light in the
mid-2020s.

The ELTs and their instruments will revolutionize all areas of
exoplanet science --- not to mention most other sub-fields of
astronomy.  Two themes summarize the advantages of atmospheric studies
with the ELTs: {\bf high angular resolution} when operating at the
diffraction limit and {\bf high spectral resolution} enabled by the
unprecedented collecting area of these large telescopes.  When using
adaptive optics (AO) to operate at the diffraction limit, angular
resolution scales inversely with telescope diameter as $\lambda/D$ and
so with sensitivity increases as $D^4$ during AO operations. Even
during seeing-limited observations (when sensitivity scales as $D^2$)
an ELT's larger aperture makes high-resolution spectroscopy
feasible on a much wider array of targets.

This brief review focuses on atmospheric characterization of
exoplanets, especially involving  measurements of the composition,
structure, dynamics, and evolution of these planets' atmospheres.  In
a sense, this work complements the recently-published broad overview
of exoplanet atmospheres \citep{crossfield:2015b} by focusing
specifically on the potential benefits that these the ELTs'
high-resolution advantages will bring to future studies in this field.
This work specifically focuses on some of the most compelling
atmospheric science cases for four qualitatively different planet
populations: highly irradiated gas giants (Sec.~\ref{sec:hj}), young,
hot giant planets (Sec.~\ref{sec:younghot}), old, cold gas giants
(Sec.~\ref{sec:oldcold}), and small planets and Earth analogs
(Sec.~\ref{sec:hab}).

\section{Instruments and Techniques}

The main obstacle preventing detailed atmospheric characterization beyond the
Solar System is the challenge of obtaining high-precision measurements
of an exoplanet mostly obscured by the glare of its bright host star.
For example, a ``hot Neptune'' and a cool T-type brown dwarf may have
comparable luminosities: yet while dozens of the latter are routinely
studied by today's ground-based telescopes \citep{mace:2013}, only a
handful of the former have been studied even after many hours of
dedicated space telescope spectroscopy
\citep{kreidberg:2014,knutson:2014a}.

Atmospheric observations seek to somehow disentangle the fainter
planetary signature from the much brighter stellar signal. This goal
is achieved in different ways for different types of planets,
typically involving high angular and/or spectral
resolution. Table~\ref{tab:inst} summarizes the basic properties of
some representative ELT instruments that might be most useful for
atmospheric characterization, along with each project's current
name for such an instrument.

\begin{figure*}[t!]
  \includegraphics[width=0.98\hsize]{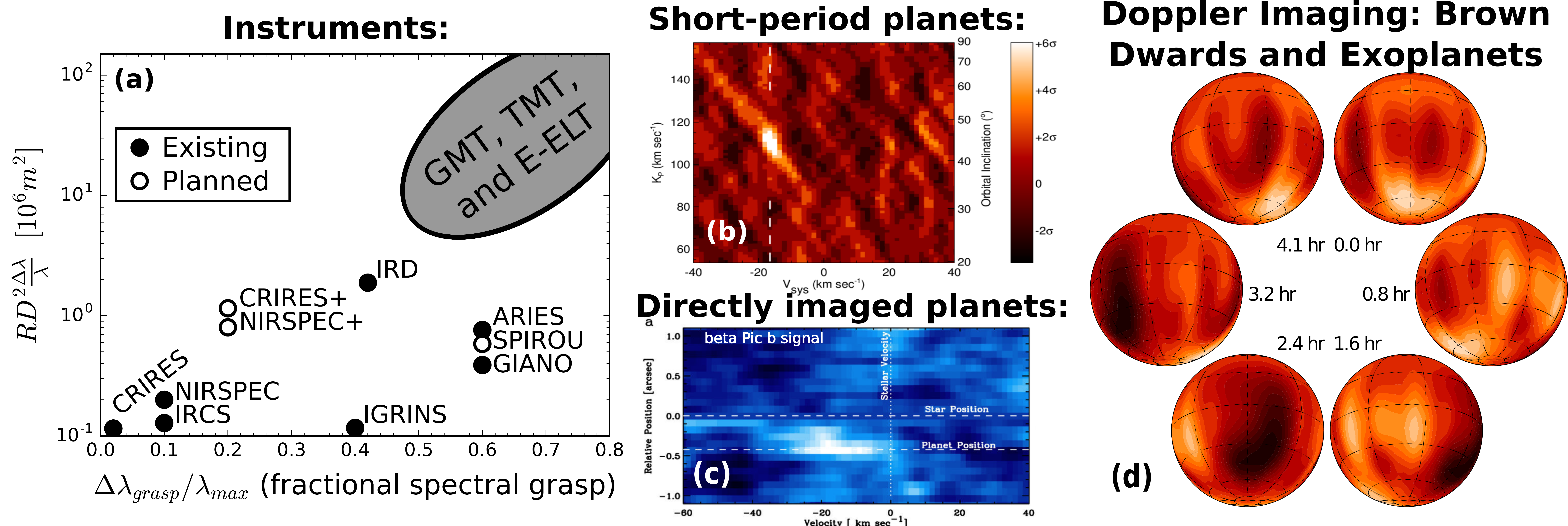}
  \caption{\footnotesize The promise of high-dispersion spectroscopy:
    {\em (a):} New and increasingly powerful infrared spectrographs
    offering broad spectral grasp and high spectral resolution enable
    transformative atmospheric characterization. These will be ideal
    for {\em (b)} measurements of atmospheric abundances, thermal
    structure, and atmospheric dynamics of short-period exoplanets
    \citep{brogi:2013}; {\em (c)} measurements of 3D orbits, angular
    momentum and rotation, and atmospheric compositions of directly
    imaged planets \citep{snellen:2014}; and {\em (d)} Doppler Imaging
    of brown dwarfs and exoplanets to create 2D global maps and
    weather movies, tracking atmospheric makeup, cloud formation and
    dissipation, and global circulation
    \citep{crossfield:2014a,crossfield:2014b,snellen:2014}.  \label{fig:map}}
\end{figure*}

For planets on shorter-period orbits ($P\lesssim20$\,d), the coherent
Doppler shift of the planet's intrinsic emission (and perhaps also
reflection) spectrum can help separate the planet from the star. The
planet-star system effectively becomes a spectroscopic binary
\citep{dekok:2013}. Hence the benefit of {\bf high spectral
  resolution}, which more effectively resolves the narrow planetary
lines (a tidally-locked hot Jupiter with Jupiter's radius and $P=1$\,d
has $v \sin i = 5$\,km~s$^{-1}$, corresponding to R\,=\,60,000).
High-resolution transit spectroscopy does not necessarily rely on a
high planetary velocity, but increased spectral resolution still helps
to separate the qualitatively different planetary and stellar spectra.
Planets on longer-period orbits experience much lower accelerations
and so Doppler-shift analyses are less effective (though, as during
transits, high spectral resolution can provide key benefits in certain
cases). More commonly, such a planet is distinguished as a separate
point source near its host star thanks to {\bf high angular
  resolution} imaging, and diffraction-limited instruments specially
designed to suppress the star's bright, scattered-light halo. When
coupled to medium- or high-resolution spectrographs, such instruments
become even more powerful.

%Broadly speaking, these instruments fall
%into two categories that take advantage of the potential for high
%spectral and/or spatial resolution: diffraction-limited cameras and/or
%integral field units (IFUs) operating in the visibla and/or NIR; and
%spectrographs (either seeing- or diffraction-limited) with
%$\lambda/\Delta \lambda \sim 10^5$ and covering optical, NIR, and/or
%MIR wavelengths.

\begin{deluxetable}{l l l l}
\tabletypesize{\footnotesize}
\tablecaption{  ELT  Instruments for Exoplanet Exploration \label{tab:inst}}
\tablewidth{0pt}
\tablehead{
\colhead{Description} & \colhead{E-ELT} & \colhead{TMT} & \colhead{GMT}
}
\startdata
$R\sim 10^5$, MIR, slit/IFU      & {\bf METIS} & MICHI & \multirow{2}{*}{GMTNIRS} \\
$R\sim 10^5$, NIR, slit      & \multirow{2}{*}{HIRES} & NIRES &  \\
$R\sim 10^5$, optical, fiber &  & HROS & {\bf G-CLEF} \\
$R\sim 3000$, NIR, IFU       & {\bf HARMONI} & {\bf IRIS} & GMTIFS \\
High-contrast imager/IFU         & EPICS & PFI/PSI & TIGER \\
\enddata
\tablenotetext{\ }{Instruments in {\bf bold} are planned for early operations.}
\end{deluxetable}

%% \renewcommand{\bottomfraction}{.5}
%% \renewcommand{\topfraction}{.5}
%% \renewcommand{\floatpagefraction}{.8}%

%\section{Telescopes, Instruments, and Capabilities}

%VLT   8.2 
%Keck  10  
%GMT   24  
%TMT   30  
%E-ELT 39

%Each ELT project necessarily has its own management structure, science
%plan, proposed instruments, and timeline.

%Types of instruments:

\section{Irradiated Gas Giants}
\label{sec:hj}

Most exoplanet atmospheres studied to date are those of highly
irradiated gas or ice giants: mostly hot Jupiters \citep[known now for over
two decades;][]{mayor:1995,charbonneau:2000} with small but
growing numbers of hot Neptunes and mini-Neptunes
\citep{butler:2004,gillon:2007,charbonneau:2009}.  These planets share
a few common characteristics: all are large enough that they must
contain substantial mass fractions of volatiles (H$_2$/He, \water,
etc.), and all have short orbital periods ($P<10$\,d) that subject the
planets to much higher levels of irradiation than seen in the Solar
System.

Models of these exotic atmospheres suggest many fascinating phenomena
that may be amenable to observation. As described elsewhere
\citep{crossfield:2015b}, these phenomena include day-to-night
temperature contrasts of hundreds to thousands of K
\citep{showman:2009}, circumplanetary wind speeds of up to several
km~s$^{-1}$ that redistributes this incident heat \citep{showman:2009},
atmospheric composition that reflects the planet's formation and
migration history \citep{oberg:2011,ciesla:2015}, unusual atmospheric
abundance patterns and metallicity enhancements 1000$\times$ or more
above the Solar composition
\citep{madhusudhan:2011b,moses:2013,fortney:2013}, temperature
inversions in the low to upper atmosphere
\citep{fortney:2008,robinson:2013}, and spatially varying abundances
and thermal structure \citep{agundez:2012,helling:2016}.

These short-period planets are best characterized via {\bf high
  spectral resolution} observations, since they orbit too near their
host stars to be resolved separately (0.1~AU/10~pc = 10~mas, or
$\sim1\lambda/D$ for an ELT). Indeed, the first atmospheric
characterization of any exoplanet's atmosphere was the detection of Na
in the hot Jupiter HD~209458b via high-resolution optical spectroscopy
during transit \citep{charbonneau:2002}. Subsequent observations have
revealed Na and/or K in the atmospheres of a growing number of hot
Jupiters \citep[for a recent summary see][]{sing:2016}. When observed at
high S/N and high spectral resolution, such observations probe alkali
abundances and the thermal structure of a planet's atmosphere
\citep{huitson:2012,heng:2015b}, as well as measuring the wind speeds on
both the dawn and twilight terminators
\citep{wyttenbach:2015,louden:2015}. After many years of searching,
the first ground-based measurement of an exoplanet's albedo has also
recently been made using high-resolution optical spectroscopy
\citep{martins:2015}. ELT instruments  will
measure all these quantities for a much wider range of planets than
the few studied in this way to date, and will provide much
higher-precision measurements of these phenomena for the most
observationally favorable systems.

Spectroscopy in the infrared is an even more powerful diagnostic of
short-period exoplanet atmospheres than are optical observations. The
planet/star contrast ratio is more favorable at longer wavelengths and
because these wavelengths host many more (and deeper) molecular
lines than do shorter wavelengths. As a result, this technique has
rapidly progressed from mere detection of molecules
\citep{snellen:2010} and constraints on cloud properties
\citep{crossfield:2011} to  measurements or upper limits on atmospheric
abundances of CO, \water, \methane, and C/O ratios; orbital
inclinations (and so absolute masses) of non-transiting planets;
thermal structure; and global rotation and winds (Fig.~\ref{fig:map}b;
\citep{brogi:2012,brogi:2013,brogi:2014,brogi:2016,rodler:2012,rodler:2013b,birkby:2013,dekok:2013,lockwood:2014}).

ELT high-resolution infrared spectroscopy will push these studies to
larger numbers of smaller, cooler planets (to date, nearly all such
studies have focused on hot Jupiters). The one exception was a
non-detection consistent with GJ~1214b's cloud-covered atmosphere
\citep{crossfield:2011,kreidberg:2014}. Fortunately some sub-Jovian,
sub-1000~K planets have at least partially cloud-free atmospheres
\citep{fraine:2014} and so new types of planets will certainly be
amenable to high-resolution spectroscopy. Furthermore, the impact of
clouds is greatly reduced when studying a planet's thermal emission
(rather than transmission) spectrum \citep{morley:2015}, providing an
alternative avenue for study. With broader wavelength coverage and
greater collecting area than existing instruments, the improved
capabilities of these new instruments (see Fig.~\ref{fig:map}a) will
allow future studies to measure precise atmospheric abundances,
measure global wind patterns and energy recirculation
\citep{kempton:2012,kempton:2014,rauscher:2014}, and (by observing at
multiple orbital phases) create longitudinally-averaged global maps of
composition, clouds, and thermal structure \citep[e.g.,][]{dekok:2014}.

\section{Young, Hot Giant Planets}
\label{sec:younghot}
The process of planet formation is a violent and, above all, energetic
process. After accretion of rocky solids into planetary cores,
considerable energy is liberated during the runaway accretion
experienced by gas giants; models of planet formation predict that for
a young, giant protoplanet achieves a luminosity as great as
$\sim10^{-4} L_\odot$ \citep{mordasini:2012b} for a few Myr.  During
this time the system exhibits an exceedingly favorable planet/star
contrast ratio. Furthermore, if gas accretion is ongoing then
traditional stellar activity indicators (e.g., H$\alpha$ emission) may
be detectable as well.  Indeed, young accreting planets have been
imaged around a few nearby systems
\citep{kraus:2011,close:2014a,quanz:2015,sallum:2015}. However, such
targets are near the limit of what can be studied using current
facilities -- with a main limitation being the $<0.1$'' separations of
these objects from their host stars. The {\bf high angular resolution}
of a diffraction-limited ELT is essential to to study a large,
representative sample of these young, accreting objects.  For example,
the study of young, giant protoplanets during formation was one of the
key science drivers behind the original instrument concept study for
the TMT's Planet Formation Instrument \citep{macintosh:2006pfi}.

Although giant planets at later ages (up to 100~Myr) are somewhat
fainter, observations (mostly photometry) have revealed considerably
more about their atmospheric composition, non-equilibrium chemistry,
luminosity \& thermal evolution, and even bulk angular momentum
\citep{konopacky:2013,bonnefoy:2014b,snellen:2014,barman:2015,morzinski:2015,skemer:2016}. All
these studies would benefit from the {\bf high angular resolution} and
increased sensitivity that an ELT's larger apertures would provide,
and many more such systems should be discovered by GAIA and ongoing
ground-based surveys by the time the ELTs begin operations.  Recent
observations reveal the even greater power of medium- to
high-resolution spectroscopy (as opposed to photometry) when
determining these planets' atmospheric properties
\citep{konopacky:2013,snellen:2014,barman:2015}. Instruments that
combine both high spatial resolution and medium-to-high spectral
resolution may therefore provide especially exciting opportunities to
expand the range of planets accessible to studies of composition,
chemistry, and clouds. Such instruments also raise the possibility of
photometric and/or spectroscopic monitoring of intrinsic variability
(weather) on these objects \citep[e.g.,][]{kostov:2013}.
 
The ELTs will also provide exciting opportunities to produce global,
two-dimensional maps via Doppler Imaging.  Fig.~\ref{fig:map}d shows
the {first 2D map of a brown dwarf} produced using this technique
\citep{crossfield:2014a}.  ELT-based high-resolution infrared
spectrographs should have the sensitivity to conduct such observations
for at least a small number of the brightest directly imaged
exoplanets \citep{snellen:2014,crossfield:2014b}. These studies will
produce global Doppler maps and weather movies of exoplanets (and many
brown dwarfs). By tracking the atmospheric dynamics and the formation,
evolution, and dissipation of clouds in these atmospheres, Doppler
imaging could provide exciting and unique insights into the 
atmospheric properties of these bodies.

\begin{figure}[tb]
\centering
\includegraphics[height=2.1in]{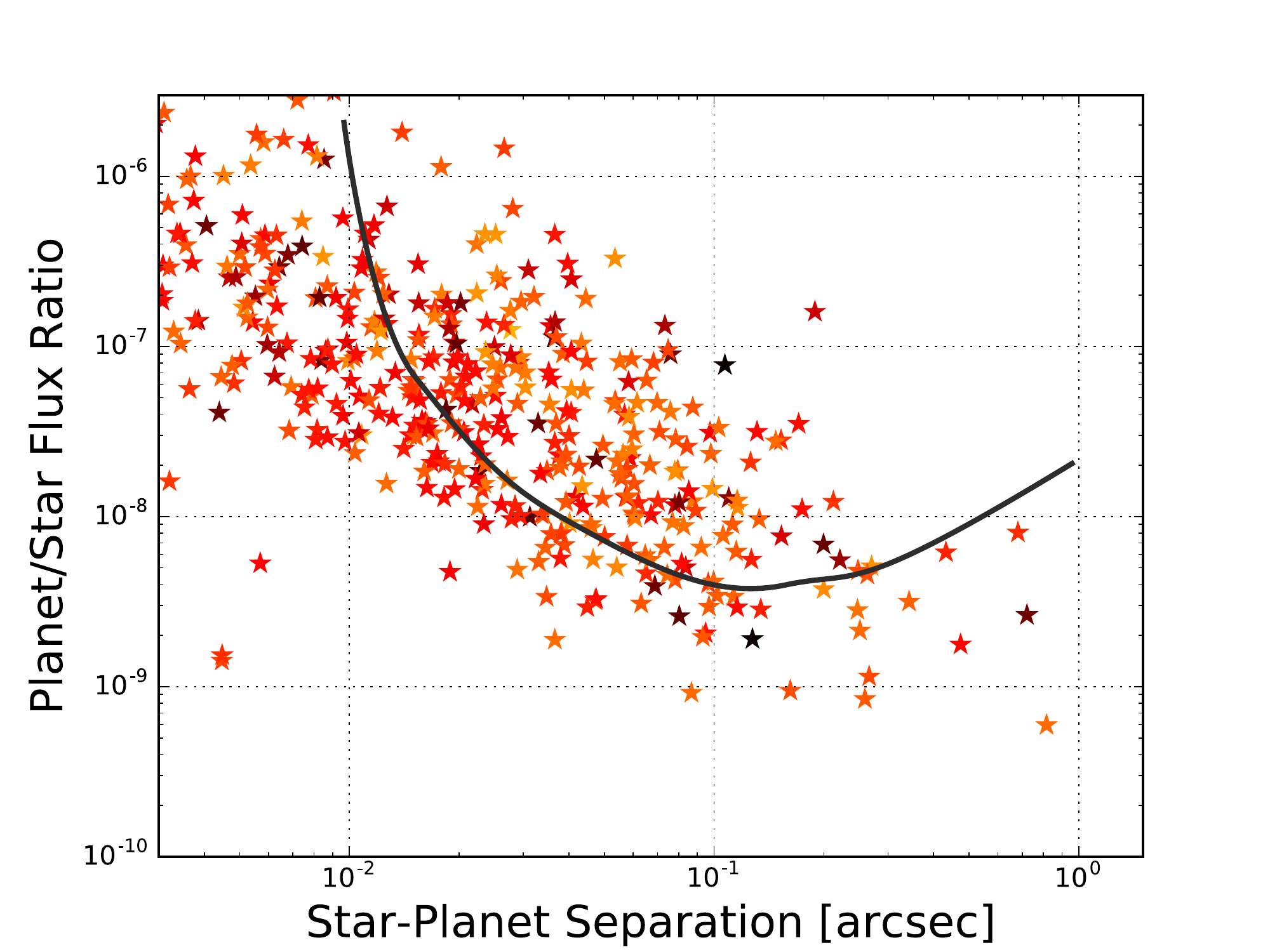}
\includegraphics[height=2.1in]{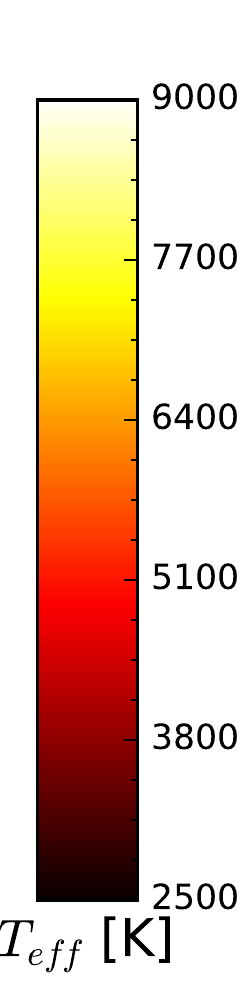}
\caption{\label{fig:knownplanets} Known RV planets potentially
  accessible to high-contrast observations in reflected
  light. Data are taken from the NASA Exoplanet Archive.  The only
  assumptions are an albedo of 0.1, $R_p/R_\oplus =
  (M_p/M_\oplus)^{0.485}$, and the radii of all planets with $R_P/R_J
  > 1.8$ and $M_p/M_J>1.8$ set to $1.2 R_J$.  The solid line is the
  approximate NIR contrast performance predicted for ELT instruments, and
  the color scale indicates $T_\textrm{eff}$. The accessible systems
  are mostly Jovian-size planets orbiting beyond 1 AU. }
\end{figure}

\section{Mature, Cold Gas Giants} 
\label{sec:oldcold}
Most high-contrast, direct imaging instruments operate in the
near-infrared and are sensitive only to the young, hot, self-luminous
giant planets on wide orbits described above.  The increased
performance (in both sensitivity and achievable planet/star contrast)
of ELT-based high-contrast instruments should allow the detection of
many old, cold, mature giants in reflected starlight around nearby
stars \citep{males:2014}.  The large numbers of ice-line gas giants detected by radial
velocity surveys \citep{mayor:2011,hasegawa:2012} indicate that there
should be substantial numbers of giant planets accessible to
high-contrast characterization in {\em reflected} light, as shown in
Fig.~\ref{fig:knownplanets}. For the first time, these
observations will allow the detailed comparison of significant numbers
of albedos, cloud and haze properties, and atmospheric abundances and chemistry
of  gas giants only marginally warmer than Jupiter and Saturn.  Such
studies will directly complement characterization of smaller numbers
of somewhat cooler giants with the WFIRST/AFTA coronagraph
\citep{marley:2014,burrows:2014b,robinson:2016}.

%\begin{figure*}[tb]
%\includegraphics[width=5in]{curves-2012-07-17_ed_v4.pdf}
%%\includegraphics[width=2.6in]{hr8799.pdf}
%\caption{\label{fig:contrast} Planet/star contrast achieved via
%  current and planned direct imaging instruments. Orange points
%  indicate a subset of detected planets (at upper right, as seen in
%  $K$ band), and Solar System planets at 10~pc (bottom, assuming
%  reflected light). Updated from \cite{mawet:2012}, as compiled by
%  \cite{crossfield:2015b}. }
%\end{figure*}

\section{Small Planets and Earth Analogs} 
\label{sec:hab}

Cooler, smaller, and more nearly Earth-like planets will remain
inaccessible to WFIRST/AFTA. Yet atmospheric characterization of
small, rocky planets lies within reach of the ELTs through {\bf high
  angular and/or spectral resolution}.  When orbiting the nearest
stars to the Sun, such planets will be accessible via high-contrast
imaging observations. Using the measured occurrence rates of small
planets around main-sequence stars measured by Kepler
\citep{howard:2012}, Monte Carlo simulations show that 10--20
short-period, sub-Jovian planets should be detectable with future ELT
instruments \citep{crossfield:2013a}. A few of the known, potentially
accessible systems plotted in Fig.~\ref{fig:knownplanets} are already
smaller than Neptune, so a preliminary target list already exists. A
fraction of these small, short-period planets could be observed in
both reflected light ($<2.5$\,$\mu$m) and thermal emission
($\sim$3--10\,$\mu$m), with the former measuring albedos and cloud
properties and the latter measuring radiometric radii via energy
balanace considerations for these planets \citep{crossfield:2013a}.
Radial velocities will measure planet masses, and will also predict
the most favorable times to observe these systems (at quadrature for
direct imaging, near opposition for Doppler-based techniques).

\begin{figure}[tb]
\centering
\includegraphics[height=2.1in]{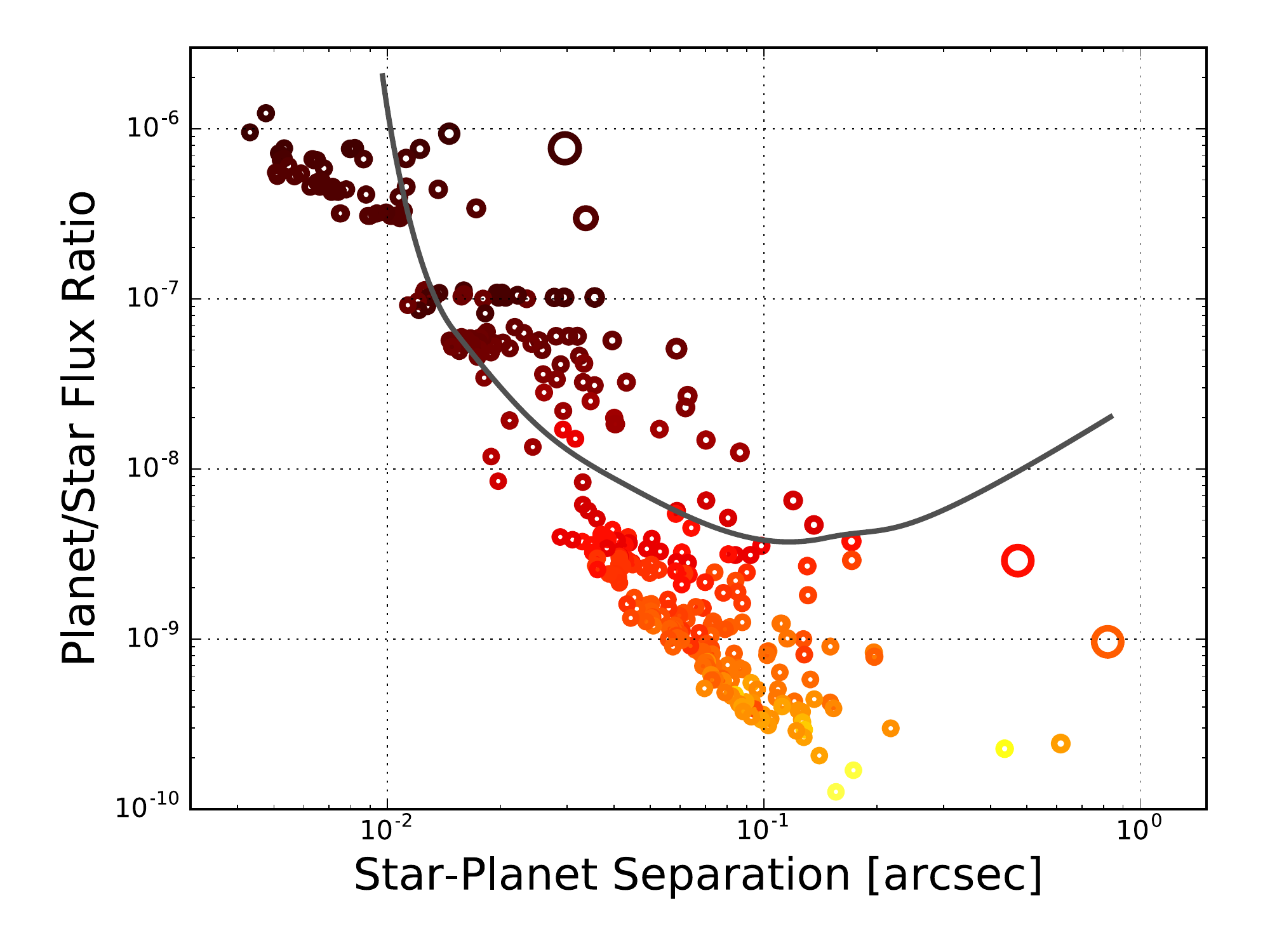}
\includegraphics[height=2.2in]{teff_colorbar.pdf}
\caption{\label{fig:habcontrast} Accessibility of hypothetical
  temperate, rocky planets to NIR-contrast observations in
  reflected light.  Each point represents a 1.25\,$R_\oplus$ planet
  receiving Earth-like irradiation, with one such planet for each star
  within 20~pc.  The solid line is the approximate NIR contrast
  performance predicted for ELT instruments. The color scale indicates
  $T_\textrm{eff}$, and point size scales inversely with distance from
  Earth. The most accessible temperate planets will orbit M1--M4
  dwarfs. }
\end{figure}

One of the most exciting and challenging long-term goals of exoplanet
studies is the atmospheric characterization of Earth analogs:
temperate, rocky planets with secondary atmospheres.  The
high-contrast instruments described above should be able to detect
such planets orbiting early-to-mid M dwarfs within 20~pc, as shown in
Fig.~\ref{fig:habcontrast}. Earth analogs orbiting earlier-type stars
are too faint relative to their host star; those around later-type
stars orbit too close to be resolved for all but the nearest systems.
Based on predicted instrument performance, Fig.~\ref{fig:habcontrast}
shows that  roughly 50 such systems could be detected if every star hosted
such a planet. Since only one in six M dwarfs hosts a rocky planet in
its Habitable Zone \citep{dressing:2015}, we should expect 5--10
temperate, rocky planets within reach of ELT {\bf high angular
  resolution} instruments.

Alternatively, the atmospheres of small, rocky planets may be studied
in transit using the same {\bf high spectral resolution} techniques
currently applied to hot Jupiters. The application of this approach to
seeking potential biosignature gases (e.g., O$_2$) has been studied
many times over the past two decades
\citep{schneider:1994,webb:2001,snellen:2013a,rodler:2014}.  The latest
(and most complete) treatment of such observations indicates that if
all visible transits are observed over a long period, one could build
up the S/N necessary for a confident detection. The timescale involved
would be of order 10 years, but only of order 45 transits would be
optimally observable and so the required observing time would be quite
manageable ($\sim$10 hr/yr). By geometric arguments, the nearest
temperate, rocky planets to the Solar System are non-transiting;
high-resolution optical spectroscopy of the type used to measure 51
Peg b's albedo \citep{martins:2015} could be used to characterize
smaller, cooler planetary atmospheres using only tens of hours of
observing time \citep{martins:2016a}.  Though these studies focus on
detecting O$_2$, the same approach could also likely characterize the
abundances of other species such as CO$_2$, \methane, \water, etc.\ on
small planets of all types and temperatures.

Finally, small and temperate exoplanets may also be studied using a
combination of both {\bf high spectral and high angular
  resolution}. This approach may be best-suited for integral field
spectrographs, though if a planet's location is well-known then
AO-fed, slit-based spectrographs may also suffice. The success of both
types of instruments when applied to known directly imaged planets
\citep{konopacky:2013,snellen:2014,barman:2015} indicates the promise
of this technique, and several studies have already considered the
applicability of this approach --- again, in the specific context of
seeking potential biosignature gases \citep{kawahara:2012,snellen:2015}.

\section{Conclusions}

The approaching era of extremely large ground-based telescopes will be
an exciting time for exoplanet science, and for atmospheric studies in
particular. In the intervening years great strides will be made with
JWST at low and medium spectral resolution, and at wavelengths from
$<1$\,$\mu$m to $\ge12$\,$\mu$m. The key advantage of the ELTs will be
their ability to deliver both {\bf high spectral resolution} and {\bf
  high angular resolution} far beyond what JWST will offer.

High-resolution spectrographs offer exciting opportunities for
measuring the composition, dynamics, structure, and cloud properties
of exoplanetary atmospheres. Fig.~\ref{fig:map} shows the science
cases soon to be enabled: global Doppler mapping of a few exoplanets
and many brown dwarfs; atmospheric composition, dynamics, and thermal
structure of short-period gas giants and sub-Jovians; rotation
measurements of directly imaged planets; and more.

High-resolution imaging and/or IFU spectroscopy will complement the
above studies by studying the composition, albedo, and cloud
properties of old, cold gas giants inaccessible to current atmospheric
characterization (see Fig.~\ref{fig:knownplanets}).  Similar
techniques should even permit the atmospheric study of smaller numbers
of rocky planets --- and (as shown in Fig.~\ref{fig:habcontrast})
perhaps even temperate, Earth-sized planets orbiting nearby M dwarfs.

\section*{Acknowledgements}

I thank the organizers of the ``20 Years of Giant Exoplanets'' and
``BashFest 2015'' meetings for allowing me to contribute to the
proceedings of both their meetings; I look forward to many future
visits to both OHP and Austin.  Thanks also to J.~Birkby, T.~Barman,
J.~Martines, and A.~Weinberger for interesting discussions about exoplanet
atmospheres and giant telescopes.  This work was performed in part
under contract with the Jet Propulsion Laboratory (JPL) funded by NASA
through the Sagan Fellowship Program executed by the NASA Exoplanet
Science Institute.

\bibliographystyle{apj_hyperref}
%\bibliographystyle{JHEP}
%\bibliography{ms.bib}

%\begin{thebibliography}{99}
%\bibitem{...}
%....
%
%\end{thebibliography}

\end{document}